\documentclass[a4paper,11pt]{article}
\usepackage{graphicx} 
\usepackage{amsmath,amssymb} 
\usepackage{bm}
\usepackage{fancyhdr}

\title{Orbit determination of space objects based on sparse optical
  data} \author{A. Milani$^{1}$, G. Tommei$^{1}$, D. Farnocchia$^{1}$,
  A. Rossi$^{2}$, T. Schildknecht$^{3}$ and R. Jehn$^{4}$
  \\ $^{1}$Dipartimento di Matematica, Universit\`a di Pisa, Pisa,
  Italy\\ $^{2}$IFAC-CNR, Firenze \& ISTI-CNR, Pisa,
  Italy\\ $^{3}$Astronomical Institute, University of Bern, Bern,
  Switzerland\\$^{4}$ESA-ESOC, Darmstadt, Germany}

\date{}

\begin{document}



\maketitle


\begin{abstract}
\footnotesize While building up a catalog of Earth orbiting objects,
if the available optical observations are sparse, not deliberate
follow ups of specific objects, no orbit determination is possible
without previous correlation of observations obtained at different
times. This correlation step is the most computationally intensive,
and becomes more and more difficult as the number of objects to be
discovered increases.  In this paper we tested two different
algorithms (and the related prototype software) recently developed to
solve the correlation problem for objects in geostationary orbit
(GEO), including the accurate orbit determination by full least
squares solutions with all six orbital elements. Because of the
presence in the GEO region of a significant subpopulation of high area
to mass objects, strongly affected by non-gravitational perturbations,
it was actually necessary to solve also for dynamical parameters
describing these effects, that is to fit between 6 and 8 free
parameters for each orbit.

The validation was based upon a set of real data, acquired from the
ESA Space Debris Telescope (ESASDT) at the Teide observatory (Canary
Islands).  We proved that it is possible to assemble a set of sparse
observations into a set of objects with orbits, starting from a sparse
time distribution of observations, which would be compatible with a
survey capable of covering the region of interest in the sky just once
per night. This could result in a significant reduction of the
requirements for a future telescope network, with respect to what
would have been required with the previously known algorithm for
correlation and orbit determination.

\end{abstract}


\section{Introduction}
\label{sec:geopop}
More than 16,000 objects with diameter larger than approximately 10 cm
are orbiting the Earth. Only about 6\% of them are operational
satellites.  All the rest is composed by different types of space
debris that now represent a serious hazard to the safe exploitation of
the circumterrestrial space.

Most of the catalogued objects reside in the Low Earth Orbit (LEO)
regime, i.e. they spend most of their life below 2000 km of
altitude. This is the region of space with the highest spatial density
of objects and where all the known collisions took place. Nonetheless
another region of space hosts a large number of spacecraft that are
crucial for our everyday life. It is the geosynchronous region,
usually defined as the part of space above about 30000 km of
altitude. This paper deals specifically with objects orbiting in this
region.

The growing risk posed by the overcrowding of the space calls for a
number of measures able in particular to minimize the risk of
collision between operational spacecraft and space debris. This
requires the accurate knowledge of the orbit of both the objects.
Currently the major effort in tracking and cataloguing the space
debris population is performed by the United States Strategic Command
(USSTRATCOM) using a large network of radar and optical sensors
located worldwide. The majority of the larger objects are catalogued
by the USSTRATCOM in the Two-Line Element (TLE) catalogue.  In this
catalogue about 16000 objects are listed along with their current
orbital parameters. The limiting size of the objects included in the
catalogue (due to limitations in sensors power and in observation and
data processing procedures) is about 5 to 10 cm below a few thousands
km of altitude and about 0.5 - 1 m in higher orbits up to the
geostationary (GEO) ones.

In particular, currently about 1000 objects, with diameter larger than
about 1 m, are classified as geosynchronous objects (mean motion
between $0.99\le$ and $1.01$ days and eccentricity not greater than
$0.01$) in the TLE cataloogue.  On the other hand, dedicated optical
campaigns from the ESA Space Debris Telescope (ESASDT) (a 1 m
telescope located on the Teide vulcano, in the Canary Islands), and
from other similar American or Russian sensors, revealed a large
number of so-called {\it uncorrelated objects}, i.e., objects not
present in the TLE catalogue. Most of these are probably the result of
a still undetermined number of explosions occurred to spacecraft and
upper stages. Dedicated optical observation campaigns were performed
to characterize the environment in this orbital region (e.g.,
\cite{schil}) for objects down to a few tens of cm.

Moreover, in recent years, a peculiar population of objects having
mean motion around 1 and high eccentricity (as high as 0.55) was
detected by the ESASDT (\cite{schild_1}). It was shown that these are
objects with very high area to mass ratio (ranging between 1 m$^2$/kg
up to 30 m$^2$/kg) whose dynamics is therefore strongly perturbed by
the solar radiation pressure that significantly affects their
eccentricity (and also their inclination) with small effects on the
total energy of the orbit and, therefore, on the semi-major axis or
mean motion (\cite{liou}). Most probably these objects are remnants of
thermal blankets or multi-layer insulation (MLI) either detached from
aging spacecraft or ejected by explosive fragmentations of old
spacecraft. It is worth noting that, from an observational point of
view, these objects represent a particularly demanding task. Their
dynamics is extremely difficult to model, due to the large influence
of the solar radiation pressure, further complicated by the unknown
and rapidly changing physical properties of the objects.  This
translates in a comparable difficulty in the determination of their
orbits.  In Sec.~\ref{sec:nongrav} the algorithm used for the orbit
determination of high area-to-mass ratio (A/M) objects will be
described.

Until recently, most of the dedicated observations have not been
devoted to cataloguing purposes and have not led to a full orbit
determination. The information obtained in the surveys made since 1999
are mainly statistical since no attempt has been made to catalog the
objects. This means that some objects may have been observed multiple
times.  From a probabilistic analysis, in (\cite{jehn}) it is pointed
out that the population of debris, brighter than visual magnitude
18.5, inferred from the ESASDT, may indeed suffer from multiple
observations. This might have lead to the over-estimation of this
particular population by a factor of about 5.

The procedures described in this paper were devised to solve this problem
and to provide effective algorithms for the building of a
European catalogue, analogous to the TLE one, foreseen in the
framework of the European Space Situation Awareness (SSA) initiative.
The SSA intends to provide Europe with an autonomous capacity 
to monitor the circumterrestrial space allowing a safe 
exploitation of this resource.

In Sec.~\ref{sec:algo} and Sec.~\ref{sec:nongrav} we briefly recall
the main features of the algorithms developed by our group in the last
years for the orbit determination of space objects. Then the dataset
used to validate the algorithms is presented.  And, finally, the
results obtained are presented and discussed.

\section[]{Algorithms}
\label{sec:algo}
Given two or more sets of observations, the main problem is how to
identify which separate sets of data belong to the same physical
object (the so-called \emph{correlation} problem). Thus the orbit
determination problem needs to be solved in two stages: first
different sets of observations need to be correlated, then an orbit
can be determined; this combined procedure is called \emph{linkage} in
the literature (see \cite{milani_recovery}).

Two different linkage methods were developed in the last few years.
The algorithms are fully described in (\cite{tommei}),
(\cite{gronchi}), (\cite{orbdet_book}), (\cite{farnocchia10}).  In
this section, for ease of reading, we will briefly recall the main
features of these algorithms, directing the reader to the above cited
papers for the full mathematical treatment.

\subsection[]{Observations and attributables}
\label{sec:att}
To understand the results presented in the following sections
some nomenclature and definitions have to be introduced.

The batches of observations which can be immediately assigned to a
single object give us a set of data that can be summarized in an {\em
  attributable}, that is a 4-dimensional vector. To compute a full
orbit, formed by 6 parameters, we need to know 2 further quantities.

Let $(\rho,\alpha,\delta) \in \mathbb R^+ \times [0,2\pi) 
\times (-\pi/2, \pi/2)$ be topocentric spherical coordinates for the position
of an Earth satellite. The angular coordinates $(\alpha,\delta)$ are
defined by a topocentric reference system that can be arbitrarily
selected.  Usually, in the applications, $\alpha$ is the right
ascension and $\delta$ the declination with respect to an equatorial
reference system (e.g., J2000). The values of range $\rho$ and range
rate $\dot\rho$ are not measured.

We shall call {\em optical attributable} a vector
\[ 
{\cal A}_{opt}=(\alpha,\delta,\dot\alpha,\dot\delta) \in [0,2\pi)
\times (\pi/2,\pi/2) \times \mathbb R^2\, ,
\]
representing the angular position and velocity of the body at a time
$t$ in the selected reference frame (for the definition of the
radar-attributable see \cite{tommei}).

Given the attributable ${\cal A}$, to define an orbit the values of
two unknowns quantities (e.g., $\rho$ and $\dot\rho$) need to be found
at the same instance in time as the attributable. These two
quantities, together with ${\cal A}$, give us a set of {\em
  attributable orbital elements}
\[
X = [\alpha,\delta,\dot\alpha,\dot\delta,\rho,\dot\rho]
\]
at a time $\bar t$, computed from $t$ taking into account the
light-time correction: $\bar t=t-\rho/c$ ($c$ being the velocity of
light). Of course the information on the observer station must be
available.


Starting from an attributable, we would like to extract sufficient
information in order to compute full preliminary orbits.

\subsection[]{Virtual debris algorithm}
\label{sec:vda}
The first algorithm developed is called the Virtual debris algorithm
and makes use of the so-called \emph{admissible region}.

The admissible region replaces the conventional confidence region as
defined in the classical orbit determination procedure. The main
requirement is that the geocentric energy of the object is negative,
so that the object is a satellite of the Earth.

\begin{figure}
  \begin{center}
  \includegraphics[width=8cm]{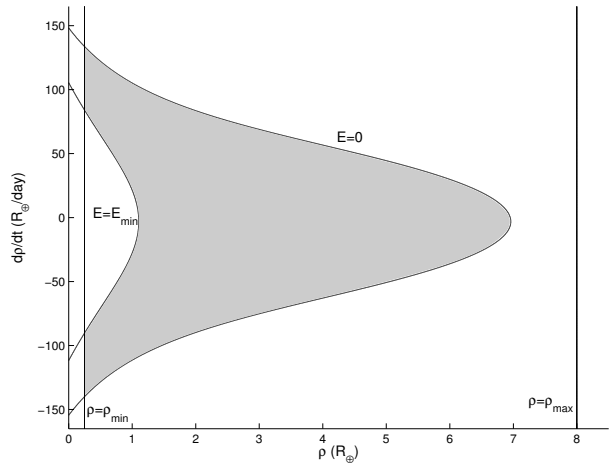}
  \end{center}
  \caption{An example of admissible region, optical case, in the
    $(\rho,\dot\rho)$ plane. The region (painted in grey) is bounded
    by two level curves of the energy, ($E=E_{min}$) and ($E=0$), and
    by the two conditions on the topocentric distance
    ($\rho=\rho_{min}$ and $\rho=\rho_{max}$).}
\label{fig:adregion_optical1}
\end{figure}

Given the geocentric position $\mathbf r$ of the debris, the
geocentric position $\mathbf q$ of the observer, and the topocentric
position $\bm\rho$ of the debris we have $\mathbf r=\bm\rho+\mathbf
q$. The energy (per unit of mass) is given by
\begin{equation}\label{energy} 
{\cal E}(\rho,\dot\rho)=\frac{1}{2}||\dot{\mathbf
  r}(\rho,\dot\rho)||^2-\frac{\mu}{||\mathbf r(\rho)||}\, ,
\end{equation}
where $\mu$ is the Earth's gravitational parameter. Then a definition of
admissible region such that only satellites of the Earth are allowed
includes the condition
\begin{equation}\label{negative_energy}
{\cal E}(\rho,\dot\rho)\leq 0 .
\end{equation}
This condition translates in a region of $(\rho,\dot\rho)$ having at
most two connected components (even if in a large number of numerical
experiments with objects in Earth orbit, we have not found examples
with two connected components.)  The admissible region needs to be
compact in order to have the possibility to sample it with a finite
number of points, thus a condition defining an inner boundary needs to
be added. The choice for the inner boundary depends upon the specific
orbit determination task: a simple method is to add constraints
$\rho_{min}\leq\rho\leq\rho_{max}$ allowing, e.g., to focus the search
of identifications to one of the three classes LEO, MEO (Medium Earth
Orbits) and GEO. Another natural choice for the inner boundary is to
take $\rho\geq h_{atm}$ where $h_{atm}$ is the thickness of a portion
of the Earth atmosphere in which a satellite cannot remain in orbit
for a significant time span. As an alternative, it is possible to
constrain the semimajor axis to be larger than
$R_\oplus+h_{atm}=r_{min}$, where $h_{atm}$ is the radius of the Earth
atmosphere.  The qualitative structure of the admissible region is
shown in Fig.~\ref{fig:adregion_optical1}.
The shaded region of Fig.~\ref{fig:adregion_optical1} can be
further constrained and reduced excluding
trajectories impacting the Earth in less than one revolution,
which means to impose that the perigee is larger than a given
value $r_{min}$ (see \cite{farnocchia10}).

Once the admissible region is defined it has to be discretized
sampling it to generate a swarm of \emph{virtual debris}.  This is
done using the Delaunay triangulation (\cite{ons1}).  The idea is to
generate a swarm of virtual debris $X_i$, corresponding to the nodes
of the admissible region of one of the two attributables, let us say
${\cal A}_1$. Then we compute, from each of the $X_i$, a prediction
${\cal A}_i$ for the epoch $t_2$, each with its covariance matrix
$\Gamma_{{\cal A}_i}$. Thus for each virtual debris $X_i$ we can
compute an attribution penalty $K_4^i$ (\cite{ons2},
\cite{orbdet_book}[Cap. 8]) and use the values as a criterion to
select some of the virtual debris to proceed to the orbit computation.

Thus the procedure is as follows: we select some maximum value
$K_{max}$ for the attribution penalty and if there are some nodes such
that $K_4^i\leq K_{max}$ we proceed to the correlation confirmation.
If this is not the case, we can try with another method, such as the
one described in Sec.~\ref{sec:kepint}.


\subsection[]{Keplerian integrals method}
\label{sec:kepint}

An alternative method to produce preliminary orbits starting from two
attributables ${\cal A}_1$, ${\cal A}_2$ of the same object at two
epoch times $t_1$ and $t_2$, was proposed for the asteroid case in
\cite{gronchi} and is based on the two-body integrals.  The method was
implemented and adapted to the space debris case
(\cite{farnocchia10}). Once more the procedure is applicable to both
optical and radar observations, but only the optical case will be
recalled here.  We assume that the orbit between $t_1$ and $t_2$ is
well approximated by a Keplerian 2-body orbit, with constant energy
${\cal E}$ and angular momentum vector $\mathbf c$:
\begin{equation}\label{keplerian_system}
\begin{cases}
{\cal E}(t_1)-{\cal E}(t_2)=0\\
\mathbf c(t_1)-\mathbf c(t_2)=0
\end{cases}
\, .
\end{equation}
Solving the system (\ref{keplerian_system}) requires a complex
analytical and numerical procedure, involving algebraic equations.
This is detailed in \cite{gronchi} and \cite{farnocchia10} and it is
not worth recalling here. Once the roots of the equation are obtained,
given all the roots which could be real, we select the positive
couples $(\rho_1,\rho_2)$ and remove the spurious ones. If the number
of remaining solutions is zero, the attributables cannot be correlated
with this method, otherwise the selected couple represents the sought
for solution.

Once a solution of (\ref{keplerian_system}) is computed the values of
attributable elements can be obtained for the epochs $\bar t_1$ and
$\bar t_2$, and they can be converted into the usual Keplerian
elements:
\[
(a_j,e_j,I_j,\Omega_j,\omega_j,\ell_j)\ ,\ j=1,2\ ,
\]
where $\ell_j$ are the mean anomalies. The first four Keplerian
elements $(a_j,e_j,I_j,\Omega_j)$ are functions of the 2-body energy
and angular momentum vectors ${\cal E}_j$, $\mathbf c_j$, and are the
same for $j=1,2$. Thus the result can be assembled in the
8-dimensional vector:
\[
  H=(V,\Phi_1,\Phi_2) \ \ ,\ \
  V=(a,e,I,\Omega) 
\]
\begin{equation}\label{vector8}
  \Phi_1=(\omega_1,\ell_1) \ ,\
  \Phi_2=(\omega_2,\ell_2) 
\end{equation}
There are compatibility conditions between $\Phi_1$ and $\Phi_2$ to be
satisfied if the two attributables belong to the same object:
\begin{equation}\label{compatibility_conditions}
\omega_1=\omega_2\ ,\ \ell_1=\ell_2+n(\bar t_1-\bar t_2)\ ,
\end{equation}
where $n=n(a)$ is the mean motion.  We cannot demand the exact
equality in the formulae above, because of various error sources,
including the uncertainty of the attributable, and the changes on the
Keplerian integrals due to the perturbations with respect to the
2-body model. Thus we need a metric to measure in an objective way the
residuals in the compatibility conditions.  The two attributables
${\cal A}_1, {\cal A}_2$ have been computed from the observations by
using a least squares fit to the individual observations, thus
$4\times 4$ covariance matrices $\Gamma_{{\cal A}_1}$ and
$\Gamma_{{\cal A}_2}$ are available; they can be used to form the
block diagonal $8\times 8$ covariance matrix for both attributables
$\Gamma_{\cal A}$. The Keplerian integral method allows to compute
explicitly the vector $H$ of (\ref{vector8}) and, by means of the
implicit function theorem, its partial derivatives, thus it is
possible by the standard covariance propagation formula
\cite{orbdet_book}[Sec. 5.5] to compute also $\Gamma_H$, the
covariance of $H$. With another transformation we can compute the
average elements $\Phi_0=(\Phi_1+\Phi_2)/2$ (as the best value for the
angular elements at time $\bar t_0=(\bar t_1+\bar t_2)/2$) and the
discrepancy $\Delta\Phi$ in the compatibility conditions
(\ref{compatibility_conditions}), and to propagate the covariance also
to this 8-dimensional vector:
\[
\Gamma_{\cal A} \Longrightarrow \Gamma_H \Longrightarrow
\Gamma_{V,\Phi_0,\Delta\Phi}\ .
\]
The above argument is a generalization of the one in \cite{gronchi},
where explicit computations are given for the optical attributables
case.

In the $8\times 8$ covariance matrix $\Gamma_{V,\Phi_0,\Delta\Phi}$,
the lower right $2\times 2$ block is the marginal covariance matrix of
$\Delta\Phi$, from which we can compute the normal matrix and the
$\chi^2$:
\[
C_{\Delta\Phi}=\Gamma^{-1}_{\Delta\Phi}\ \ , \ \ \chi^2_{\Delta\Phi}=
\Delta\Phi\cdot C_{\Delta\Phi}\,\Delta\Phi\ ,
\]
which can be used as control, that is the discrepancy in the
compatibility conditions is consistent with the observation error and
the correlation between the two attributables is considered possible
only if $\chi^2_{\Delta\Phi}\leq \chi^2_{max}$.

The upper left $6\times 6$ block is the covariance matrix of the
preliminary orbit, that is of the orbital elements set $(V, \Phi_0)$
(at epoch $\bar t_0$). Although this preliminary orbit is just a
2-body solution, it has an uncertainty estimate, arising from the
(supposedly known) statistical properties of the observational
errors. This estimate neglects the influence of perturbations, such as
the spherical harmonics of the Earth gravity field, the lunisolar
differential attraction and the non-gravitational perturbations;
nevertheless, if the time span $\bar t_2-\bar t_1$ is short, the
covariance obtained above can be a useful approximation.  
Recently the method was generalized, including the effect due to the
non-spherical shape of the Earth (\cite{farnocchia10}), thus allowing
its application also to objects in LEO. On the other hand since the
present paper deals only with high Earth orbit data, where the effect
of $J_2$ on the angular momentum of the objects is strongly reduced by
the distance from the center of the Earth, all the analysis presented
in this paper was performed without the inclusion of the $J_2$ effect.

Note that there are some cases in which the Keplerian integrals method
can not be applied. We have to avoid the condition $(\mathbf
q_1\times\hat{\bm\rho}_1)\times(\mathbf q_2\times\hat{\bm\rho}_2)=0$,
where $\mathbf q_1$ and $\mathbf q_2$ are the observer geocentric
positions at the instants $t_1$ and $t_2$. This can happen when:
\begin{itemize}
\item $\mathbf q_1$ is parallel to $\hat{\bm\rho}_1$, i.e., the
  observation at time $t_1$ is done at the observer zenith;
\item $\mathbf q_2$ is parallel to $\hat{\bm\rho}_2$, i.e., the
  observation at time $t_2$ is done at the observer zenith;
\item $\mathbf q_1$, $\mathbf q_2$, $\hat{\bm\rho}_1$ and
  $\hat{\bm\rho}_2$ are coplanar. This case arises
  whenever a geostationary object is observed from the same station at
  the same hour of distinct nights.
\end{itemize}
As it is normal, the mathematical singularity is surrounded by a
neighborhood in which the method is possible for zero error (both zero
observational error and zero rounding off in the computation), but is
not applicable in practice due to the limited numerical accuracy;
e.g., this method fails even for non-geostationary, nearly
geosynchronous orbits with hours of observations over different nights
differing by only a few minutes each night. Note that in an
observing strategy optimized for the use of this method,
this occurrence can be easily avoided. 

\subsection{Correlation confirmation}
\label{sec:correl_confirm}

The multiple orbits obtained from the solutions of the algebraic
problem are just preliminary orbits, solution of a 2-body
approximation (as in the classical methods of Laplace and Gauss).
They have to be replaced by least squares orbits, with a dynamical
model including all the relevant perturbations.

Even after confirmation by least squares fit, it might still be the
case that some linkages with just two attributables can be
\emph{false}, that is the two attributables might belong to different
objects. This is confirmed by the tests with real data reported in
\cite{tommei_esoc} for the virtual debris method and in
\cite{milani_esoc} for the Keplerian integrals method.  Thus every
linkage of two attributables needs to be confirmed by correlating a
third attributable.

The process of looking for a third attributable which can also be
correlated to the other two is called \emph{attribution}
(\cite{milani_recovery,attrib}). From the available 2-attributable
orbit with covariance we predict the attributable ${\cal A}_P$ at the
time $t_3$ of the third attributable, and compare with ${\cal A}_3$
computed from the third set of observations. Both ${\cal A}_P$ and
${\cal A}_3$ come with a covariance matrix, we can compute the
$\chi^2$ of the difference and use it as a test. For the attributions
passing this test we proceed to the differential corrections.
The procedure is recursive, that is we can use the 3-attributable
orbit to search for attribution of a fourth attributable, and so
on. This generates a very large number of many-attributable orbits,
but there are many duplications, corresponding to adding them in a
different order.

A specific procedure, called \emph{correlation management} is used to
remove duplicates (e.g., $A=B=C$ and $A=C=B$) and inferior
correlations (e.g., $A=B=C$ is superior to both $A=B$ and to $C=D$,
thus both are removed). The output catalog after this process is
called normalized. In the process, we may try to merge two
correlations with some attributables in common, by computing a common
orbit (\cite{ons2}).


Due to the characteristics of the two methods briefly outlined in this
Section, it can be noticed that the two algorithms have different
ranges of application.  The virtual debris algorithm should be applied
to short time intervals between observed arcs, less than one orbital
period or at most a few orbital periods. The Keplerian integrals
method, thanks to the constancy of the integrals of the 2-body problem
even over significant time intervals, can be used for longer time
spans, spanning several orbital periods. On the other hand, it is near
to a singularity for very short time spans and in some other
near-resonance conditions, such as observations of a geosynchronous
orbits at the same hour in different nights.
We conclude that each method should be used in the cases in which it
is most suitable as will be illustrated in Sec.~\ref{sec:results}.  

\section[]{Non gravitational perturbation model}
\label{sec:nongrav}
The solar radiation pressure represents the largest non-gravitational
perturbation acting on a spacecraft in high Earth orbit.
As detailed in (\cite{mi-no-fa}), the solar radiation pressure mainly 
accounts for periodic perturbations in the eccentricity $e$ and inclination $i$ 
of the orbit. On the other hand, whenever the orbit is such that the
satellite periodically enters the shadow of the Earth
(as in the case of the GEO satellites), the eclipses 
have an important perturbative effect on the
orbit, because there could be a secular effect on semimajor axis $a$,
thus an accumulated along track displacement quadratic in time.
The situation becomes worse in the case of the high $A/M$ objects
where the solar radiation pressure can become the dominant perturbative 
term beyond the spherical Earth approximation for $A/M\simeq 10$ m$^2$/kg.
Therefore the perturbations can result in significant changes in $a$
and/or in very large values of $e$ and $i$ (\cite{valk-anselmo}).
Moreover for this kind of objects very little is known about their
physical properties thus preventing an effective modelling of
the non-gravitational perturbations affecting them.

Other non-gravitational effects can contribute with a secular
perturbation in $a$, see e.g. \cite[Chap. 14]{orbdet_book}, including
the so-called Yarkovsky effect, which is the result of a systematic
anisotropic emission of radiation due to uneven external surface
temperature, and indirect radiation pressure, due to radiation
reflected and/or re-emitted by the Earth. These effects are smaller
than the main component of radiation pressure in terms of the
instantaneous value of the force, by a factor typically somewhere
between a few parts in $1000$ and a few parts in $100$. Still, they
can be the dominant source of perturbation in the satellite position
after a number of orbital periods, while the main source of short term
perturbations remains, in almost all cases, the main anti-Sun
component.

For the above reasons an adaptive non-gravitational perturbations
semi-empirical model, with the following properties was developed:
\begin{itemize}
\item 
For observed arcs either of total duration $\leq 0.01$ days,
or with less than 3 tracklets, we use no non-gravitational
perturbation model, thus we solve for each set of correlated
observations for only 6 orbital elements.
\item 
For observed arcs with at least 3 tracklets and total duration
$>0.01$ days we use a model with direct radiation pressure, only the
anti-Sun component, and with a free $A/M$
parameter\footnote{Actually, the parameter incorporates the
so-called \textit{reflection coefficient}, which cannot be
separately determined and is anyway close to 1.}, thus we solve
for at least 7 parameters.
\item 
For observed arcs with at least 4 tracklets and total duration
$>2$ days we use a model with an additional secular along track term
giving quadratically accumulated along track displacement, with a
free multiplicative parameter with the dimension of $A/M$ (to ease
comparison with the other term) thus we solve for 8 parameters.
\end{itemize}
The direct radiation pressure model includes a model for eclipses
(with penumbra), thus it already includes some quadratic
perturbations when the orbit is subject to eclipses.

The controls used to activate the more complex models take into
account not just the time span but also the amount of observational
data available in order to preserve the
over-determined nature of the least square fit. E.g., if we were to use
2 tracklets in an 8-parameters fit, there would be only 8 equations in
8 unknown. In particular, the Keplerian Integrals method of
Section~\ref{sec:kepint} has shown a good capability of finding an
approximating 2-body solution even for cases in which the orbit is
moderately perturbed, such as a large $A/M$ case over several days.
If we were to attempt a fit with non-gravitational perturbations with
the initial correlation, that is still with 2 tracklets, a 7-parameter
orbit would be very weakly determined and instabilities of the
differential corrections iterations could result in abandoning many good
correlations.

Note that the semi-empirical models, such as this one, contain terms
which are not in a one to one correspondence with physical effects. E.g.,
the along track term may represent a Yarkovsky effect as well as
resulting in secular perturbations in $a$.

Although the parameters are fitted, one caution is important: when
using a semi-empirical parameter such as $A/M$, we need to constrain
the values which can be determined within a physically meaningful
range. We are currently using $[-1, +200]$ as the control range for
the $A/M$ coefficient (in m$^2$/kg) for direct radiation pressure, and
$[-1, +2]$ for the one of the along track force.

\section[]{Observations and survey strategies}
\label{sec:obs}
For the purpose of this study it was decided to use existing data from
observations performed at ESA's 1-meter Space Debris Telescope
(ESASDT). The data stems from surveys and so-called follow-up
observations of the year 2007. The former were optimized to search for
small-size debris in the GEO region and the geostationary transfer
orbit region (GTO), with the main objective to derive statistical
information. Follow-up observations, on the other hand, are used to
maintain a catalogue of debris objects to allow for detailed analysis
of physical characteristics, e.g. by acquiring multi-color photometry,
spectra, etc. It is important to notice that the surveys were not
designed in a way to serve as a test for a ``survey only'' catalogue
build-up and maintenance strategy. As a consequence the resulting
observations were not intended to serve as test data for orbit
determination or tracklet correlation algorithms. Survey strategies
optimized to build-up and maintain a catalogue of objects without the
need of explicit follow-up observations are feasible, but should
obviously be designed in close connection with the tracklet
correlation and orbit determination algorithms.

Space debris observations at the ESASDT are organized as monthly
observation campaigns consisting of about 10 to 14 nights centered on
New Moon.  Generally, there are three types of observations performed:
\begin{itemize}
	\item \textit{GEO surveys}, with a search area optimized for GEO
		orbits with $0-20^\circ$ inclination. The tracking during the
		exposure (so-called 'blind tracking') is optimized for object in
		GEO.

	\item \textit{GTO surveys}, with a search area optimized for GTO
		orbits with $0-20^\circ$ inclination (Ariane GTO launches). The
		blind tracking during the exposure is optimized for objects in
		GTO.

	\item \textit{Follow-up observation} for a subset of the objects
		discovered in surveys (maintenance of a catalogue of debris
		objects). The total arcs covered by follow-up observations range
		from a few hours up to many months.
\end{itemize}

Table~\ref{tab:tab_ch} gives an overview of all the ESA GEO and GTO
campaigns from January 2007 until December 2007. The terms
``correlated'' and ``uncorrelated'' refer to objects/tracklets for
which a corresponding catalogue object could or could not be
identified, respectively. The identification procedure, or
``correlation procedure'', is based on comparing the observed orbital
elements and the observed position in longitude and latitude of the
object at the observation epoch with the corresponding data from the
catalogue. We used the unclassified part of the USSTRATCOM catalogue
as our reference.
\begin{table}
\begin{center}
\caption{ESA GEO and GTO Campaigns}
\label{tab:tab_ch}
\begin{tabular}{@{}lc}
\hline
&  Jan - Dec 2007\\
&  GEO/GTO\\
\hline
Frames & $56000$ \\
Scanned Area & $7600$ deg$^2$ \\
Total Observation Time & $81$ nights / $461$ h \\
GTO / Follow-up & $180$ h/$193$ h \\
Correlated tracklets & $483$\\
Correlated objects & $241$ \\
Uncorrelated tracklets & $618$ \\
\hline
\end{tabular}
\end{center}
\end{table}

The data set for the test of the algorithms was provided by the
Astronomical Institute of the University of Bern (AIUB).  It contained
the tracklets of all correlated and uncorrelated ``objects'' from the
2007 GEO and GTO surveys, as well, as the tracklets from all follow-up
observations. For this data independent information about tracklets
belonging to one and the same object, at least for the correlated
objects and the objects which were followed-up intentionally, are
available.

The data set contains $3177$ tracklets, among them
\begin{itemize}

\item $977$ uncorrelated tracklets,

\item $747$ correlated tracklets of $349$ correlated objects
  ("correlated" = correlated with USSTRATCOM TLE catalogue),

\item $1453$ tracklets from intentional follow-up observations of
  $240$ objects.
\end{itemize}

The uncorrelated and the correlated tracklets were found in the GEO
and GTO surveys, but also during follow-up observations instead or in
addition to tracklets of objects to be followed up.  The surveys
covered the GEO region rather homogeneously but were
not optimized to re-observe objects, e.g. from night to night.  Based
on results by \cite{jehn} and \cite{schild_statanal} these
$977$ uncorrelated tracklets could belong to $300 - 500$ objects.

The tracklets of the objects which were intentionally followed-up have
very particular characteristics, which are non-typical for survey
observations and thus worth mentioning. These objects belong to an
AIUB-internal catalogue of small-size debris in GEO- or GTO-like
orbits. The catalogue is biased towards objects with high area-to-mass
ratio due to deliberate selection. For a newly detected object the
standard procedure consists of acquiring 1-4 follow-up observations
during the night of discovery, resulting in arcs of 0.5-5
hours. Additional 1-2 follow-up observations are then performed during
the nights following the discovery, eventually followed by regular
observations every month. It is, again, worth noticing that the
temporal distribution of these follow-up observations does not at all
represent typical space debris survey or surveillance (SSA) scenarios.

\begin{figure}
  \begin{center}
    \includegraphics[width=8cm]{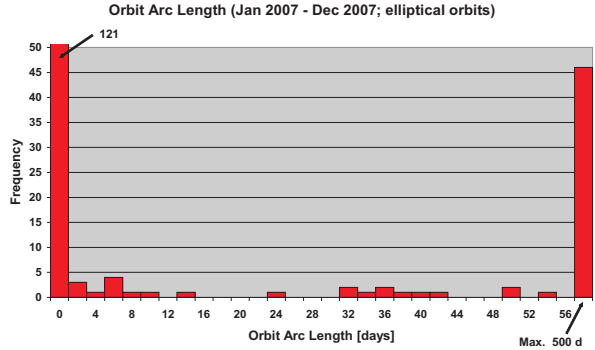}
  \end{center}
  \caption{Histogram of the arc length for the objects which were
    intentionally followed-up during the 2007 campaigns.}
  \label{fig:arcl_JanDec2007}
\end{figure}

The available arc length for the majority of the objects which were
followed-up during the 2007 campaigns is less than one day (see
Fig.~\ref{fig:arcl_JanDec2007}). For 46 objects, however, an arc
length of more than 57 days is available.

The tracklet data was provided in the form of so-called {\bf ``tracklet
files''} of the Pan-STARRS Data Exchange Format (DES).  The DES is
described by a complex document, which introduces the necessary
concepts, fixes one standard terminology, defines the data types with
an object-oriented style, assigns formats and procedures for
exporting/importing all the data types.

In these files, observations pertaining to the same tracklet are
identified by a unique ``tracklet identifier''. The assignment of
individual observations to a tracklet is (by definition) done by the
"observer" as it is intimately related with the survey and the object
detection algorithms. A typical survey will, though, not provide any
information about ``objects'', i.e. about the mutual correlation of
tracklets. However, if such information is available, it may be coded
in the so-called ``secret names''. This information (discriminating
between uncorrelated tracklets, correlated objects or follow-up
observation of correlated objects) is not to be used in the test
phase, but it is stored in order to allow a final comparison with the
``ground truth''.

\section[]{Results from a one year experiment}
\label{sec:results}

The new algorithms and the related software described in
Sec.~\ref{sec:algo} were applied to the data set described in the
previous section.  The purpose was to show that they are adequate for
a future catalog buildup activity by ESA, e.g., in the context of the
SSA initiative. Thus we selected a time interval long enough that we
can presume a future SSA survey would have observed all target objects
within such a period, and short enough to allow for accurate orbit
determinations with our semi-empirical non-gravitational perturbations
model. We selected the lunation as a kind of natural time unit for
observations. The tracklets of objects observed several times within
one lunation should be correlated. On the contrary, objects observed
only once per lunation may not be correlated, because this is well
beyond the SSA specifications.

\subsection[]{The test on one year of data}
\label{sec:oyd}

As explained in Sec.~\ref{sec:obs} the data set contained three
classes of tracklets: the ones correlated by attribution to TLE
objects, the ones correlated by AIUB (in most cases, by targeted
follow up), and the ones for which no correlation was previously
known.

The database of tracklets was split in 12 lunations. The algorithms
described in Sec. \ref{sec:algo} were applied to each lunation
separately.  The correlations within each lunation are normalized,
thus there are no duplicates, using the correlation management
procedure (see Sec. \ref{sec:correl_confirm}).

Out of 3177 input tracklets, 1503 were correlated, 1674 left
uncorrelated. Of course we have no way to know how many should have
been correlated, that is how many physically distinct objects are
there: in particular, objects re-observed at intervals longer than 10
days have escaped correlation, because we did not try to perform the
first step when the time span between two tracklets exceeds 10
days. As already pointed out in Sec. \ref{sec:obs}, the observations
were not scheduled to allow for orbit determination of all the
objects, but only for some of them, in particular the uncorrelated
objects, which were of interest as candidate high A/M cases.

\subsection[]{The global orbit catalog}
\label{sec:goc}
Joining the orbits computed in each lunation, we obtained 202
correlations with a good orbit and more than two tracklets. This process
might generate duplications of orbits for the same object. In fact if
two orbits for the same object are computed in different lunations it is
not always possible to correlate them, especially if the two lunations
are not consecutive. We plan to investigate the issue of duplications
in this catalogue in the future.

\begin{figure}
  \begin{center}
\includegraphics[width=8cm]{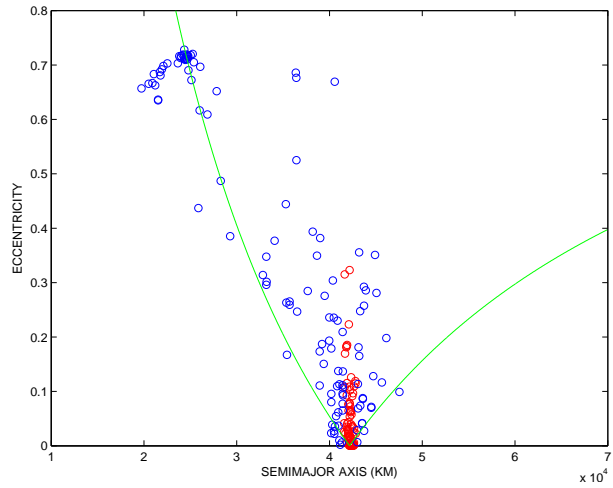}
  \end{center}
\caption{Distribution in semimajor-axis vs. eccentricity of the
  computed orbits. Red circles indicate objects with semimajor-axis
  between 41464 and 42864 km, i.e. nearly geostationary. Blue circles
  indicate all the other orbits. The green lines bound orbits crossing
  the GEO radius at apogee (left curve) or at perigee (right curve).}
\label{fig:ae}
\end{figure}

\begin{figure}
  \begin{center}
\includegraphics[width=8cm]{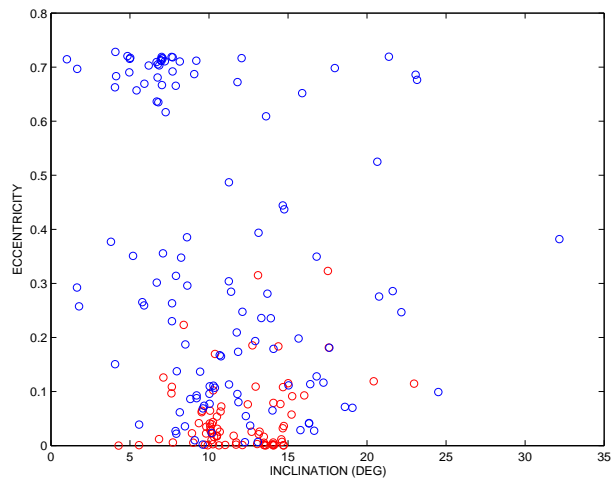}
  \end{center}
\caption{Distribution in inclination vs. eccentricity of
  the computed orbits. Red circles indicate objects with
  semimajor-axis between 41464 and 42864 km, i.e. nearly
  geostationary. Blue circles indicate all the other orbits.}
\label{fig:ie}
\end{figure}

\begin{figure}
  \begin{center}
\includegraphics[width=8cm]{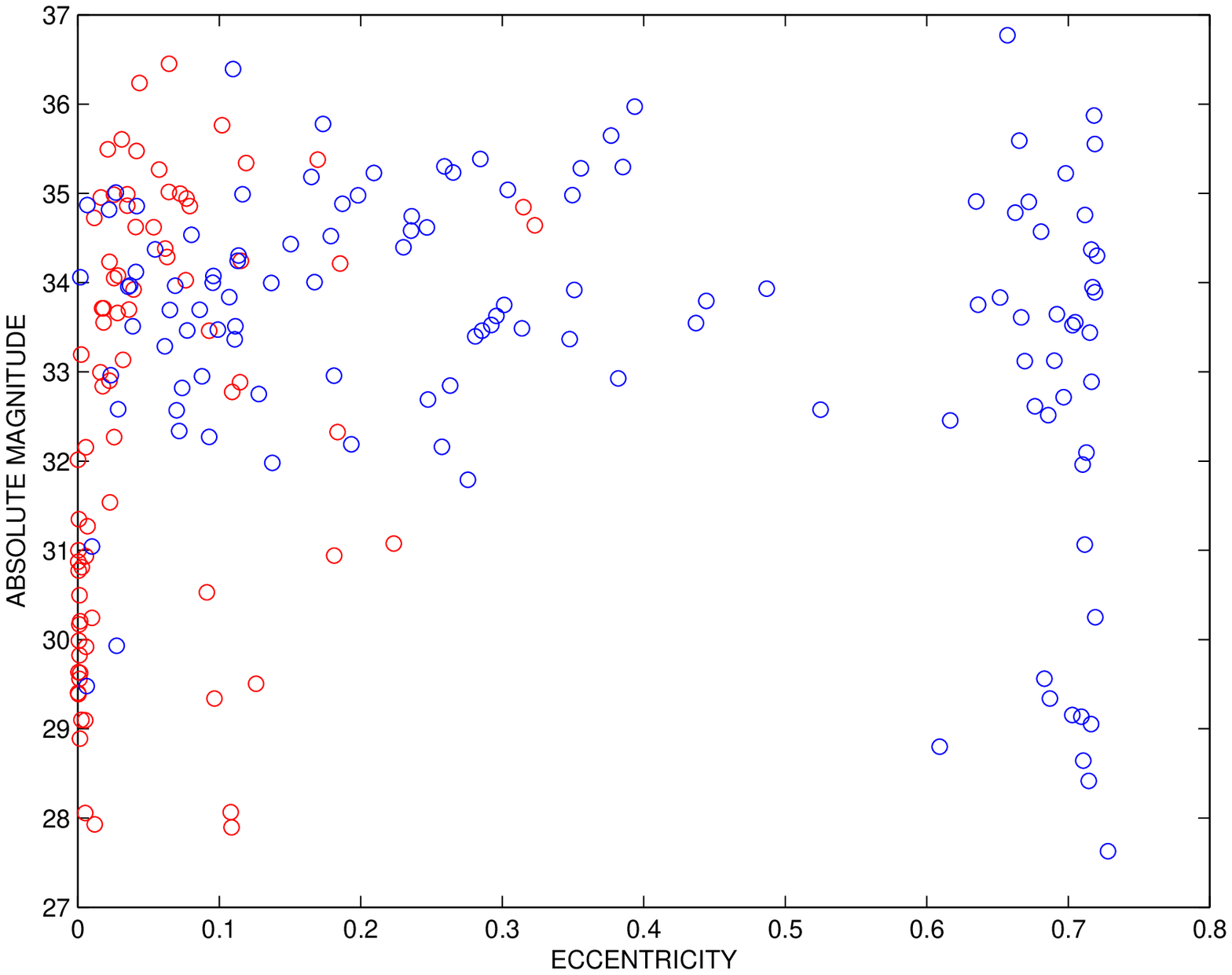}
  \end{center}
\caption{Distribution in eccentricity vs. absolute magnitude of the
  computed orbits. Red circles indicate objects with semimajor-axis
  between 41464 and 42864 km, i.e. nearly geostationary. Blue circles
  indicate all the other orbits.}
\label{fig:eh}
\end{figure}

\begin{figure}
  \begin{center}
\includegraphics[width=8cm]{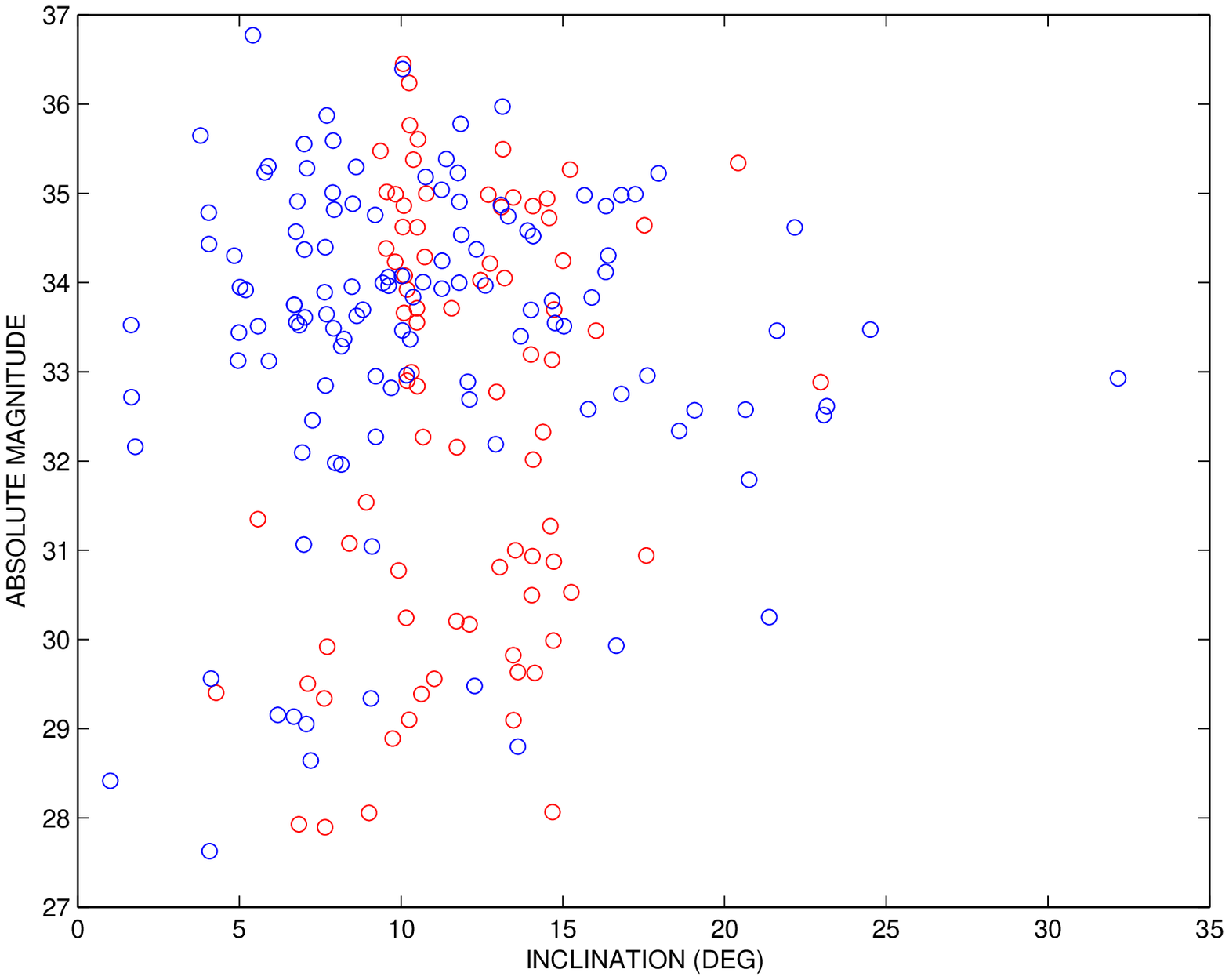}
  \end{center}
\caption{Distribution in inclination vs. absolute magnitude of the
  computed orbits. Red circles indicate objects with semimajor-axis
  between 41464 and 42864 km, i.e. nearly geostationary. Blue circles
  indicate all the other orbits.}
\label{fig:ih}
\end{figure}

Figures \ref{fig:ae}--\ref{fig:ih} show the distribution of
the computed orbits in terms of orbital elements and absolute
magnitude.


The orbits in the $(a,e)$ plane (Fig.~\ref{fig:ae})
show a concentration of objects with semimajor-axis close the
geostationary radius, including some with high eccentricity.  Some of
these latter objects have a high value of the A/M parameter, as
described in Sec. \ref{sec:geopop}.  In the upper left corner the
objects in GTO can be found with $e\simeq 0.7$. Fig.~\ref{fig:ie}
shows the same orbits in the $(I,e)$ plane. 

Fig.~\ref{fig:ie} shows an apparent lack of
really geostationary orbits, with low $e$ and $I$: actually there is
only one orbit with $e < 0.01$ and $I < 5^\circ$. This is due to the
fact that the survey conducted by the ESASDT in 2007 had the purpose of
discovering new objects, and the geostationary objects are mostly
active satellites, whose orbits and ephemerides are known. Thus the
fields of view were on purpose avoiding the geostationary line of
Fig.~\ref{fig:ae}.


Figs.~\ref{fig:eh} and \ref{fig:ih} show the distribution of
eccentricity/inclination versus intrinsic luminosity of the objects,
the latter described in the absolute magnitude scale. Unfortunately it
is not easy to convert an absolute magnitude into a size, because of
the wide range of albedo values and also because of irregular
shapes. However, if we could assume albedo 0.1 and a spherical shape,
we would get a diameter ranging between 10 m and $\simeq$ 30 cm for
the correlated objects. Thus the largest objects should be satellites
(at low $e$) and rocket stages (near GTO), the smallest are certainly
debris.

The existence of objects with high $e$ and also $I$ was already well
known, what is interesting is that some of these have a quite large
cross section. To understand the dynamics of such objects is a
challenge, which requires advanced models and a good data set of both
astrometry and photometry.

\subsection[]{Determination of non-gravitational parameters}
\label{sec:det-nongrav}
As previously pointed out in Sec.~\ref{sec:nongrav} it is important to
succeed in determining the perturbations due to non-gravitational
effects.  As described in Sec.~\ref{sec:algo}, the algorithms
were modified in order to handle this task. This implies not just to
have a non-gravitational perturbation model in the orbit propagator,
but to apply the adaptive model progressively as the correlations
build up, with the semi-empirical parameters gradually added to the list
of variables to be solved.

\begin{figure}
  \begin{center}
\includegraphics[width=8cm]{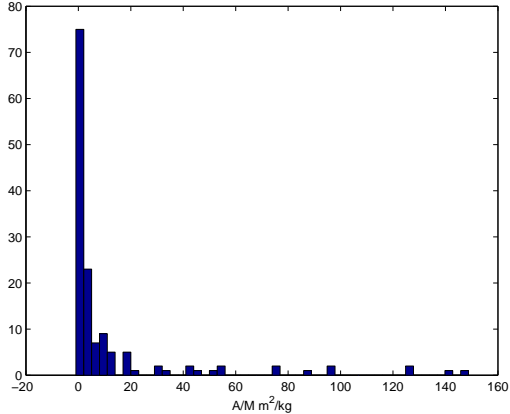}
  \end{center}
\caption{Histograms showing the number of objects with a significant
  value for A/M parameter.}
\label{fig:ng1_hist}
\end{figure}

\begin{figure}
  \begin{center}
\includegraphics[width=8cm]{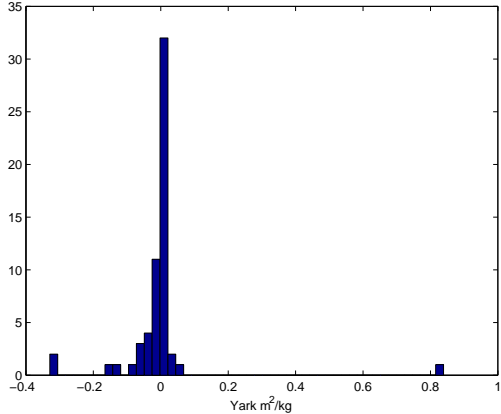}
  \end{center}
\caption{Histograms showing the number of objects with a significant
  value for the along track perturbation Yark parameter.}
\label{fig:ng2_hist}
\end{figure}

In the Figs.~\ref{fig:ng1_hist} and \ref{fig:ng2_hist} the
distribution of the values of non-gravitational parameters computed
for a subsample of the objects displayed in Fig.~\ref{fig:ae} are
shown. In particular, the A/M parameter was computed for 143 objects
and for 59 objects also the along track perturbation (called Yark
parameter) was determined. Note that the name of the latter parameter
is just suggestive, we cannot discriminate between a true Yarkovsky
effect (due to thermal emission) from an effect of direct radiation
pressure on a complex shape debris, as discussed in
Sec.~\ref{sec:nongrav}.

Whereas the bulk of the objects lies in the first histogram bin, a
significant fraction of them belong to the so-called large area to
mass ratio population.  Note that a few objects display a huge value
of A/M ($>100$ m$^2$/kg) and Yark (absolute value $>0.5$
m$^2$/kg). However, these cases typically have a large uncertainty,
possibly due to the too short time span between the correlated
tracklets.


The problem is that we did not have any ``ground truth'' to compare
our results on non-gravitational perturbations; for this we would need
to have a catalog with orbits and non-gravitational parameters from
other sources. On the other hand the objects on which to perform such
a comparison should be carefully selected, among those with the best
determination not just of the orbit but also of the semi-empirical
parameters. These ``good cases'' might require fitting more than one
month of data. This problem will also need to be further investigated.

\subsection[]{Assessment of the results}
\label{sec:asses}
Once a catalog of orbit is obtained it would be important to be able
to judge the performance of the algorithms and the reliability of the
catalogue itself. In the present analysis no absolute ``ground truth''
(that is an orbit catalog used as input for the data simulation) was
available to validate the catalog. Nonetheless a meningful comparison,
giving an indication of the validity of the procedure, was possible by
comparing with the correlation results obtained by the group that
produced the data set itself.  This implies that it is not always
possible, in case of a discrepancy between the two catalogs, to decide
``who is right''. Fortunately, this was not necessary. The goal was to
show that the new algorithms allow to obtain substantially the same
results obtained by the AIUB group, without having access to the
scheduling information. Namely, if an uncorrelated object has been the
target of deliberate follow up, the AIUB group had the correlation
information a priori (and the same information could be obtained for a
correlated object, just by comparing with the ephemerides). On the
other hand the present analysis did not use any a priori information.

To make an in depth study, we selected the two lunations which
included the largest number of tracklets, namely the first and the
second one.  In Tables \ref{tab:lun1} and \ref{tab:lun2} we show a
summary of the results obtained.

The meaning of the table columns is the following:
\begin{itemize}
\item \emph{equal} indicates the cases in which the new algorithms obtained the same
  correlation reported by AIUB;
\item \emph{larger} indicates the cases in which the new algorithms added some additional
  tracklets (marked NS) to those considered by AIUB in their correlation;
\item \emph{new} indicates correlations not found by AIUB, i.e., orbits
  computed by the new algorithms using just NS tracklets;
\item \emph{smaller} indicates the cases in which the new
  algorithms got a correlation  using a subset of the tracklets used by AIUB;
\item \emph{missed} indicates the cases in which the new algorithms did not get the
  correlation reported by AIUB;
\item \emph{mixed} indicates the cases in which the new algorithms obtained a correlation
  using a partly different set of tracklets with respect to AIUB. That
  is the new algorithms got a correlation using some (but not all) of the tracklets
  exploited by AIUB and, at the same time, added some NS tracklets.
\end{itemize}

\begin{table}
\begin{center}
\caption{Summary of the comparison with AIUB for the first lunation. 
  Between parenthesis we highlight the number of occurrences where we
  identified the reason for the smaller or missed correlations with an
  observation strategy not optimized for our algorithms. See text for details.}
\label{tab:lun1}
\begin{tabular}{@{}ccccccc}
\hline
Number & Equal & Larger & New & Smaller & Missed & Mixed\\
of Tracklets &  &  &  &  &  & \\
\hline
16 & - & - & - & - & - & 1\\
10-11 & 1 & - & - & 1 & - & -\\
7-8 & 7 & 1 & 1 & - & - & -\\
4-6 & 7 & 3 & - & 1 & 1 (1) & -\\
3 & 4 & 3 & 2 & 3 (3) & 1 (1) & -\\
\hline
\end{tabular}
\end{center}
\end{table}

\begin{table}
\begin{center}
\caption{Summary of the comparison with AIUB for the second lunation. 
  Between parenthesis we highlight the number of occurrences where we
  identified the reason for the smaller or missed correlations with an
  observation strategy not optimized for our algorithms. See text for details.}
\label{tab:lun2}
\begin{tabular}{@{}ccccccc}
\hline
Number & Equal & Larger & New & Smaller & Missed & Mixed\\
of Tracklets & & & & & & \\
\hline
11-12 & - & 3 & - & - & - & 1\\
7-9 & 3 & - & - & 3 (1) & - & 1\\
4-6 & 12 & 5 & - & 8 (4) & 7 (6) & -\\
3 & 10 & 1 & 3 & 3 (2) & 6 (4) & 1\\
\hline
\end{tabular}
\end{center}
\end{table}

A deeper analysis of the underlying reasons for the smaller and missed
correlations shows that some of them could be traced back to the
observation strategy. As pointed out several times, the observation
strategy adopted by AIUB to obtain the data used in this study was not
intended for the exploitation of the algorithms described in
Sec.~\ref{sec:algo}. In particular the requirement of avoiding the
singularities described in Sec. \ref{sec:kepint} was not considered,
because the very existence of such a problem was not known at the
time.

As already discussed, the two algorithms have a
limiting time span (different for the two methods) between consecutive
tracklets above which a correlation is unlikely to be found. An
observation strategy optimized for the use of these algorithms should
take into account this requirement, but for the same reason above,
this was not the case for the AIUB data used in this study.  The cases
in which we were able to attribute the smaller or missed correlations
to the observational strategy are highlighted in the tables with the
number written between parenthesis: these cases includes both the
observations at the same hour in the night and the observations
separated by a time interval exceeding 5-6 days.

The cases of 2-tracklet correlations were deemed not reliable. As a
matter of fact the typical RMS in the semimajor axis for this orbits
were thousands of km for observations taken in the same
night. Therefore the probability of being true if a longer time span
was available is judged to be very low. A comparison among the
2-tracklet correlations proposed by AIUB and the new method shows a
very large fraction of disagreement. Although we would like to confirm
this with a specific test, the indication is that 2-tracklet
correlations are to be considered as an intermediate data product, not
a result, that is they are accepted only if and when it is possible to
confirm them with the correlation of a third tracklet.

The results just discussed show a good agreement to the ones obtained
by AIUB. The number of cases of ``superior'' results (columns
\emph{larger} and \emph{new}) compensate the ``inferior'' results
(columns \emph{smaller} and \emph{lost}), especially if the cases in
which the observing strategy was unsuitable are discounted.

Thus we showed that it is possible to build up a catalog from scratch,
without any prior correlation information. This \emph{catalog buildup}
phase is necessarily the first phase of a new program such as SSA,
because correlation information is not available, or available only
for a comparatively small subset of the target population of the new
survey.

Moreover, we showed that the presence of non-gravitational
perturbations, whose parameters are not known a priori and can be
quite large, does not increase the difficulty of the initial catalog
buildup.  The determination of some non-gravitational perturbation
parameters can be done simultaneously with the correlation and orbit
determination procedure. To achieve this goal a suitable observing
strategy should be used: in particular for the geosynchronous belt one
tracklet per night is enough, but ``equal hour'' singularities and too
long time intervals should be avoided.

Of course to obtain the result of building up a large catalog of
satellites and space debris, down to sizes smaller than the ones for
which orbits are now available, requires the mobilization of
appropriate resources. These include sensors more powerful than the
current experimental ESASDT (in particular with a larger field of view),
and adequate software, such as a scheduler with the capability of
taking into account the requirements from orbit determination, and a
fully tested correlation and orbit determination software which could
be based upon the prototype we have developed.

\section[]{Conclusions}
\label{sec:conclusion}
As stated by ESA, ``the European Space Situational Awareness (SSA)
Programme serves the strategic aims of the European Space Policy (ESP)
by supporting the independent capacity to securely, sustainably and
safely operate Europe's critical space infrastructure''.

In the next few years the SSA initiative will enter its definition and
practical implementation phase.  One of the goals of the SSA programme
is to provide a European catalogue of Earth orbiting objects similar
to the American TLE. This goal requires the realization of a Space
Surveillance Network of radar and optical sensors able to detect and
track a large number of objects.

Whereas the definition of the network is still in progress, it is
clear that the availability of efficient methods for orbit
determination is of paramount importance in improving the efficiency
of the network. It is worth stressing that an efficient and
computationally intensive orbit determination procedure can act in a
twofold way in the definition of the network. From one side, given a
certain network design, it allows to reach more ambitious goals in
terms of cataloguing performances, e.g. allowing the cataloguing of
objects with lower diameter limit or the cataloguing of more elusive
objects such as the high A/M objects.  It must be noted at this stage
that the size limit within the TLE catalogue is dictated not only by
sensor limitations, but also by limits in the handling and computer
processing of the observational data.  On the other side, given the
preliminary requirements of a surveillance network (e.g., in terms of
the minimum size of the objects to be catalogued), the adoption of an
efficient orbit determination method allows significant savings in the
design of the sensors.

In this paper it was shown how the methods described in
Sec.~\ref{sec:algo} allowed the determination of six-parameters orbits
from a standard dataset of optical observations.  No a-priori
information nor simplified assumptions (such as circular orbits) were
required and the observation strategy was completely independent from
the design of the methods and not optimized for their use.  Even the
most demanding cases of high A/M objects were successfully treated.

The results of this study clearly show that the methods described in
Sec.~\ref{sec:algo} can represent an important tool in the SSA data
processing.  In \cite{farnocchia10} an extension of the Keplerian
integrals method, including the $J_2$ perturbation, was presented thus
allowing the correct treatment of objects in LEO.  The application of
the method to LEO optical and radar data will be tested in the near
future.

\section*{Acknowledgments}
This study was performed under ESA/ESOC Contract n. 21280/07/D/CS.

\end{document}